\documentstyle[proceedings,psfig]{crckapb} 

\def \mdot {\rm{M}$_\odot$~yr$^{-1}$}
\def \kms{km~$\rm{s}^{-1}$}

\begin{opening}
\title{HYDRODYNAMIC COLLIMATION OF YSO JETS}

\author{Adam Frank} 
\institute{Department of Physics and Astronomy,\\
Bausch and Lomb Building, University of Rochester,\\
Rochester, NY 14627-0171, USA}
\author{Garrelt Mellema} 
\institute{Stockholm Observatory, S-13336 Saltsj{\"o}baden, Sweden}

\end{opening}

\runningtitle{HYDRODYNAMIC COLLIMATION}

\begin{document}

\maketitle

\begin{abstract}
We present the results of numerical hydrodynamic models for the collimation
of outflows from young stellar objects. We show that the presence of a
toroidal environment can lead to efficient formation of jets and bipolar outflows
from initially uncollimated central winds. The interaction 
between the wind and the
environment leads to two types of collimation, one which is dominated by
radiative cooling effects, and one which works when cooling
is less efficient. We describe the two types of jets as they appear in
the simulations and we suggest a description for the long term evolution
of these structures in more realistic time-dependent wind sources.
\end{abstract}

\section{Introduction}
The physical processes behind the origin of the subjects of this conference,
HH jets, clearly still requires a lot of study.  Observations are, however,
taking us closer to the young stars and it seems well established that for
low mass stars the only way to generate the observed momentum and energy in
the jets and outflows is to have some form of MHD process accelerate material
into a wind.  Beyond this realization it is still not clear what type of MHD
process is at work: is it the interaction of stellar magnetosphere with the
inner edge of the accretion disk (the so-called X-wind model,
\citeauthor{Shuea94}~\citeyear{Shuea94}); or a centrifugally driven wind
coming off the inner parts of the accretion disk \cite{Kon89,Pud91}? It
should also be realized that for massive stars the situation is quite
different; there the driving force behind the wind may be the stellar
radiation.  It is also not clear if all MHD models can produce well
collimated jets on the observed lengthscales, something which was also
recently pointed out in a numerical study by
\citeauthor{Romanovaea96}~\shortcite{Romanovaea96}.

Based on these concerns we propose to separate the issues of wind
acceleration and jet and bipolar outflow collimation.  We believe such a
distinction is useful as it allows us to address issues such as the relation
between jets and outflows in a focused way and allows the possibility that
the acceleration processes may be different in low and high mass stars.

In this work we study the collimating properties of the environment. Our
approach is related to the earlier work on hydrodynamic collimation, such as
DeLaval nozzles \cite{Kon82,RaCan89}. Such models have fallen out of favor
because of length scale requirements \cite{KoRu93} and stability
considerations \cite{KoMc92}.  We have found however that these objections do
not hold when considering the time-dependent evolution of these types of
flow.

The background to our present study is the high degree of collimation found
when studying the interaction of a stellar wind with a toroidal environment
in the context of the formation of aspherical Planetary Nebulae (PNe)
\cite{Ickeea92}. The mechanism has been called ``Shock-Focused Inertial
Confinement'' (SFIC).  In the SFIC mechanism it is the inertia of a toroidal
environment rather than its thermal pressure, which produces a bipolar {\it
wind-blown bubble}.  In our studies of bipolar PNe formation we found the
bubble's wind shock (which decelerates the wind) takes on an aspherical,
prolate geometry \cite{Eic82,Ickeea92}.  The radially streaming central wind
strikes this prolate shock obliquely focusing it towards the polar axis and
initiating jet collimation.  Other effects such as instabilities along the
walls of the bubble, help to maintain the collimation of the shocked wind
flow.

This mechanism was first applied to the case of young stars by
\citeauthor{FrNor94}~\shortcite{FrNor94}, and more recently by
\citeauthor{FrMel96}~\shortcite{FrMel96} (henceforth Paper 1) and
\citeauthor{MelFr97}~\shortcite{MelFr97} (henceforth Paper 2). A related
study is the one by \citeauthor{PeEi95}~\shortcite{PeEi95}.

\section{Numerical Method}

To study the collimating effects of the environment we choose the case where
the central wind is maximally uncollimated, i.e.\ perfectly spherical.  This
can be considered the worse case scenario for MHD winds since most mechanisms
will results in some degree of focusing towards the axis.  The central
protostellar wind is fixed in an inner sphere of cells on the computational
grid. The relevant input parameters are simply the mass loss rate ${\dot
M_{\rm w}}$ and velocity $V_{\rm w}$ in the wind.

For the environment we choose a density distribution with a toroidal
shape. Such toroidal density distributions are theoretically expected for the
collapse of a rotating cloud \cite{TSC84}, a flattened filament
\cite{Hartmea96}, or a magnetized cloud \cite{LiShu96}. These are admittedly
all axi-symmetric models, and it would be interesting to see how this holds
under more general conditions.  \citeauthor{LucRoch97}~\shortcite{LucRoch97}
present some observational evidence for toroidal environments of proto-stars.

\begin{figure}
  \begin{center} \leavevmode \psfig{file=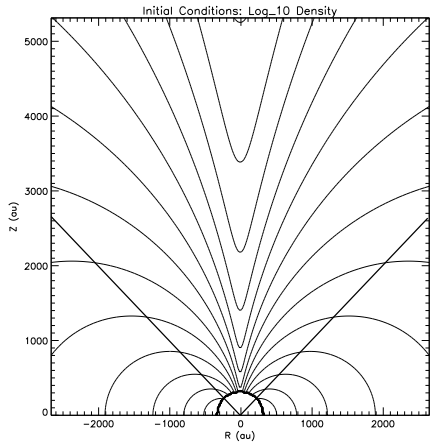,width=6cm,angle=0}
    \caption{Initial density distribution.  Shown are the $\log_{\rm 10}$
    contours of density from Eq.~\ref{rhoin} with an equator to pole contrast
    $q = 70$.  The two solid lines show the angle at which $\rho = .5
    ~\rho_{\rm max} = .5 ~\rho(90^{\rm o})$.  These occur at $\theta \sim
    45^{\rm o}$ making the opening angle of the density distribution $\sim
    90^{\rm o}$}\label{densfig}
 \end{center}
\end{figure}

The actual form of the toroid we use is given by
\begin{equation}
\rho(R,\theta) = {{\dot M}_a \over 4\pi R^2} ({2GM \over R})^{- {1 \over 2}}
{\lbrace 1 - {\zeta \over 6}{\lbrack 13P_{\rm 2} ( \cos (\theta)) -1
\rbrack}\rbrace}
\label{rhoin}
\end{equation}
in which $R$ is the spherical radius (we will use $r$ to denote the
cylindrical radius). In Eq.~\ref{rhoin} ${\dot M}_{\rm a}$ is the accretion
mass loss rate and M is the mass of the star.  Equation \ref{rhoin} is a
modified form of Eq.~96 from \citeauthor{TSC84}~\shortcite{TSC84} (originally
derived by \citeauthor{Ulrich76}~\citeyear{Ulrich76}).  We use it here
because produces the required toroidal geometry as well having the $R^{\rm
-{3 \over 2}}$ radial dependence, appropriate to a freely falling envelope.
The parameter $\zeta$ determines the flattening of the cloud, which we prefer
to quantify with the density contrast $q$, the ratio between the density at
the equator and at the pole.  We note that this density distribution is
actually very similar to the solution found by
\citeauthor{LiShu96}~\shortcite{LiShu96} for a magnetized cloud.  The shape
of this density distribution is shown in Fig.~\ref{densfig}, which serves to
illustrate that the opening angle of the toroid is actually quite wide. A
more detailed description of the initial conditions can be found in Paper 1.

The evolution of this system is followed by solving numerically for the
well-known Euler equations, including a source term in the energy equation to
account for radiative losses.  The numerical method used to solve these
equations is based on the Total Variation Diminishing (TVD) method of
\citeauthor{harten83}~\shortcite{harten83} as implemented by
\citeauthor{Ryuet95}~\shortcite{Ryuet95}.  For cooling we used the standard
coronal cooling curve of \citeauthor{DalMc72}~\shortcite{DalMc72}.  Details
of the methods and the implementation of cooling can be found in Paper 2.

\begin{figure}
  \begin{center} \leavevmode
    \psfig{file=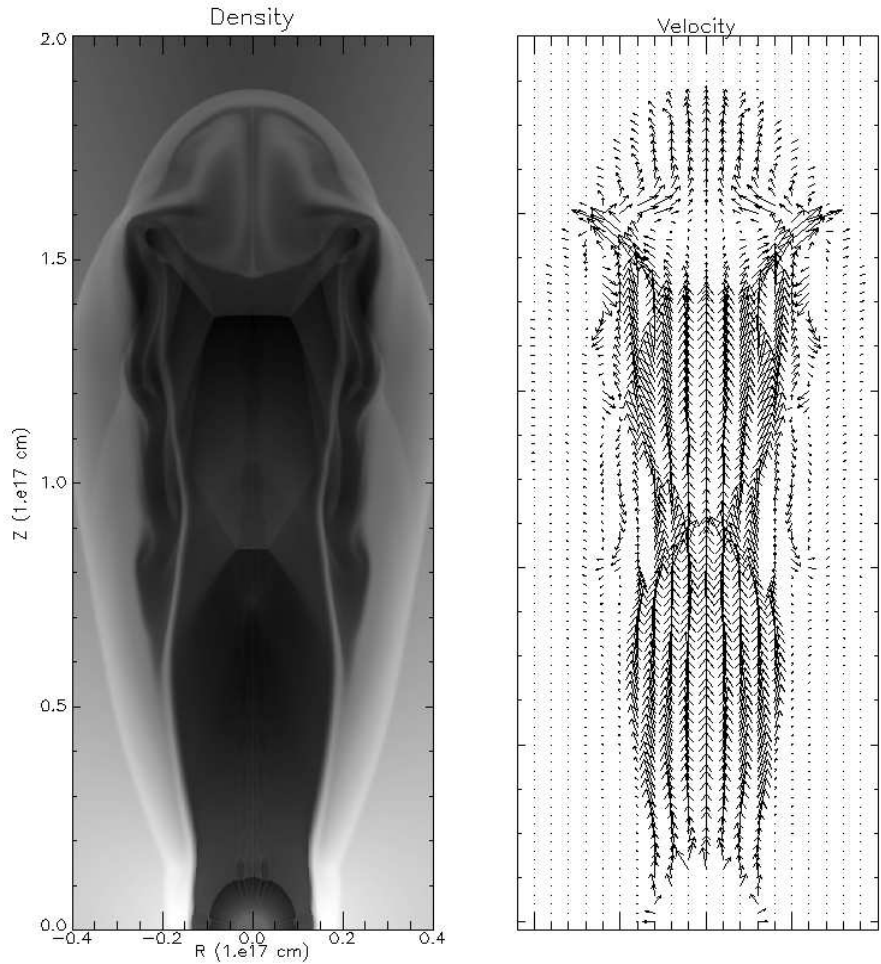,width=12cm,angle=0} \caption{Density
    and velocity for a non-cooling model.  Shown are a gray scale map of
    $\log_{\rm 10} (\rho)$ and a map of velocity vector field for
    model A after 1035 years of evolution.  In the density map dark
    (light) shades correspond to low (high) densities.  In the
    velocity field map vectors in the inner, freely expanding wind
    zone have not been plotted.  Thus the first ``shell'' of vectors
    maps out the wind shock.}
    \label{ad_logdens}
  \end{center}
\end{figure}

\section{Results without cooling}

In Paper~1 we presented the results of simulations without cooling. Although
this is not a realistic assumption in most cases, we considered it to be a
good first step. The fact that some of the features of the non-cooling jets
also appear in the simulations with cooling (see below), justifies this
approach.

Figure~\ref{ad_logdens} shows a typical result from Paper~1. The outflow and
environment parameters are: ${\dot M}_{\rm w}=10^{-7}$~\mdot,
$V_{w}=200$~\kms, ${\dot M}_{\rm a}=10^{-5}$~\mdot, $q=70$.

The collimated flow that forms has all the usual features expected for
gaseous jets: bow and jet shocks; turbulent cocoons; crossing shocks and
internal Mach disks. In Paper 1 we demonstrated that the flow pattern has
characteristics of both a supersonic jet and a wind-blown bubble, something
which may be important for understanding the connection between jets and
molecular outflows.  These simulations also showed that strong collimation is
principally achieved by the refraction of flow vectors across the aspherical
wind shock. Even a small degree of asphericity in the wind shock (a ratio
between equatorial and polar shock positions smaller than 0.8) is sufficient
to produce strong flow focusing (cf.~\citeauthor{Icke88}~\citeyear{Icke88}).

\begin{figure}
  \begin{center} \leavevmode \psfig{file=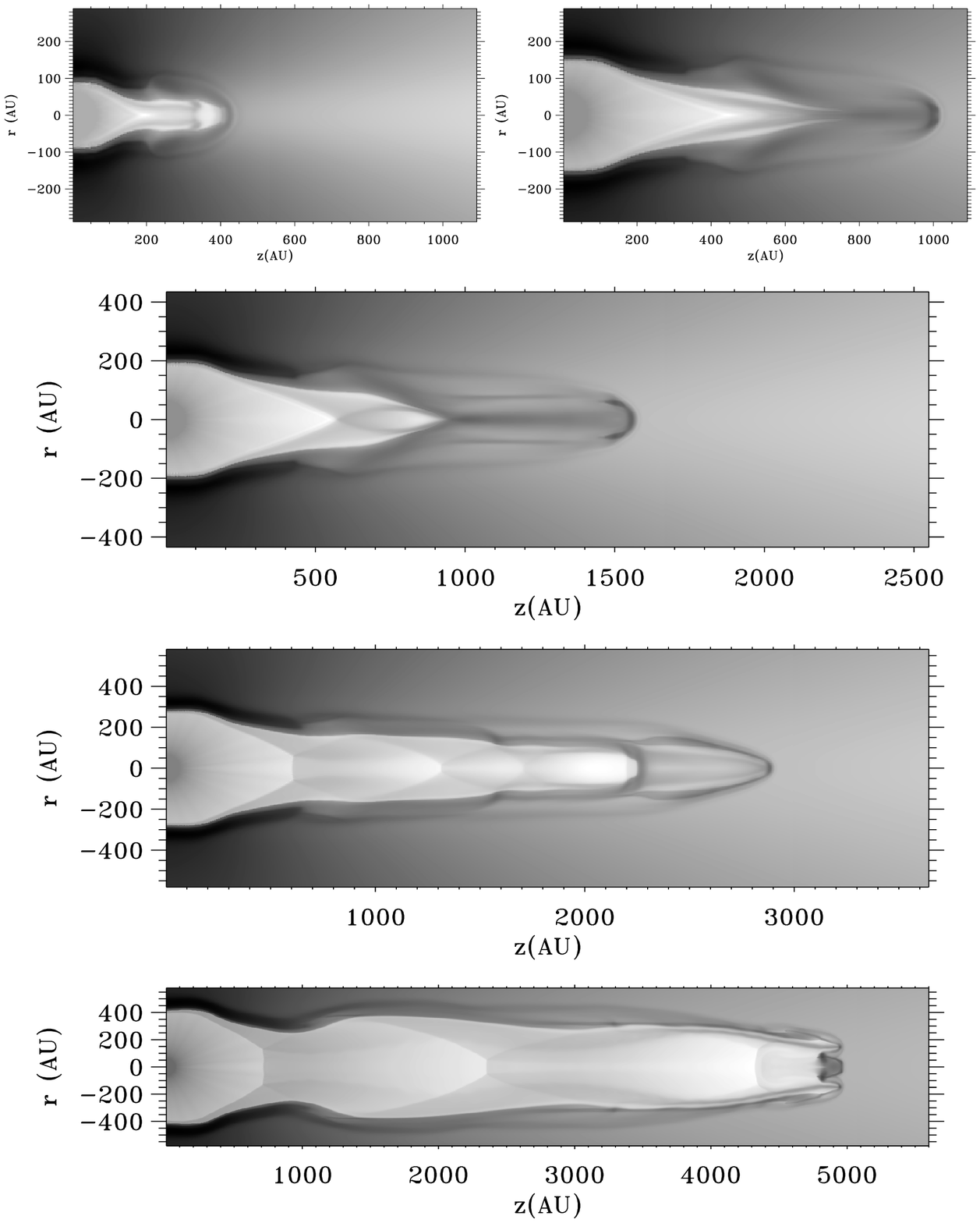,width=11cm,angle=0}
    \caption{Density evolution for a cooling simulation. The grey scales show
    $\log_{\rm 10}$ contours for times $t=$25, 62, 87, 137, 225~years. The
    darker shades are higher density. One sees the initial radiative phase,
    the development of the `cool jet' and the development of the `hot jet'.}
    \label{rad_logdens} \end{center}
\end{figure}

\section{Results with cooling}

In Paper~2 we presented the results of simulations with radiative cooling
included. Although the cooling is only implemented in a simplified way (by
using a coronal cooling curve) this should still recover the most important
dynamical effects of the cooling.

A typical sequence can be seen in Fig.~\ref{rad_logdens}.  In this simulation
the outflow has a velocity of 350~\kms\ and a mass loss rate of
$10^{-7}$~\mdot. The environment has a pole to equator contrast of 70, and an
accretion rate ${\dot M}_{\rm a}=10^{-6}$~\mdot.  We find that the evolution
of the flow proceeds in two phases. First a dense cool jet forms
($t=62$~years), followed at later times by a lower density, hot jet
($t>100$~years). This second, `hot jet' is similar to the jets found in the
non-cooling calculations of Paper~1.

The formation of this `cool jet' early in the evolution is an interesting new
effect seen in these radiative simulations. It forms when the bubble is still
mostly radiative and its formation mechanism appears to be the same as that
of the ``converging conical flows'' originally proposed by
\citeauthor{CanRod80}~\shortcite{CanRod80}, and later studied analytically
and numerically by \citeauthor{CanTenRoz88}~\shortcite{CanTenRoz88} and
\citeauthor{TenCanRoz88}~\shortcite{TenCanRoz88}. When the thermal energy
produced at the wind shock is completely lost to radiation the post-shock
flow collapses and the only path available for the shocked wind is to slide
along the contact discontinuity towards the poles of the bubble, leading to
converging conical flows.  The collision of fluid parcels at the poles
produces a shock wave where the transverse velocity components are lost and a
narrow well collimated beam or jet of dense, cool gas is formed.

In general one can say that hydrodynamic collimation is more effective when
cooling is included, which given that most of the sideward expansion is due
to thermal pressure is not too surprising. Another effect that helps in
achieving better collimation is that due to radiative losses the wind shock
is much more aspherical, leading to very efficient focusing. In fact even
during the hot jet phase the focusing is so strong that at some positions the
flow stays supersonic across the wind shock.

The duration of the cool jet phase depends very much on the efficiency of the
cooling, and hence on the velocity of the outflow and the density of the wind
and the environment. In deeply embedded objects which have higher densities
than we used in the model presented here, the cool jet will last much longer,
and a hot jet may never form.

\section{The Effect of Wind Variability} 
The results from the previous section demonstrate that strong collimation can
be achieved from a purely hydrodynamic interaction between winds and
protostellar environments.  But to apply this model to young stars we must
account for the time scales inherent to YSO jets and molecular outflows
(assuming that jets are connected to the outflows,
\citeauthor{Chernea94}~\citeyear{Chernea94}).  Recent deep exposure images of
HH jets such as HH34 \cite{BalLev94} and HH46/47 \cite{Heathea96} reveal
multiple bow shocks that imply jet lifetimes of many thousands of years or
more. In addition, the molecular outflows have dynamical lifetimes on the
order of $10^3$ to $10^5$ years \cite{Bach96}.

Because our bubbles are not in pressure equilibrium with the environment,
they will continue to grow in the lateral direction, leading to wide jets at
later times. In order to account for the observed jet widths, this expansion
has to be restricted. An attractive means for stopping this equatorial growth
comes directly from the observations.  As was mentioned above many HH jets
show clear signs of variations in velocity along the jet beams.  In the most
extreme case the jet appears to be temporarily shut off, which explains the
multiple bow-shock structures.  In the HH34 superjet
\citeauthor{BalLev94}~\shortcite{BalLev94} find a periodicity in the beam
with a period between 300 and 900 years.  In addition smaller scale
variations with periods $\tau < 100$ years are seen in many HH jet beams
(Morse, private communication).  It is reasonable that the velocity
variations in the jet beam are a measure of velocity variations in the source
material of the jet.  There might also be links with FU Orionis outbursts or
related events, which occur on similar time scales as the ones mentioned
above (see Robbins Bell, this volume)

One can therefore envisage a scenario where the wind-blown bubble is driven
by a time-variable wind.  In our simulations we did not include the effect of
either gravity or the inward directed accretion flow.  Both of these effects
will decelerate the flow and constrain the bubble, particularly near the
equator where the bubble radius is small so that the accretion velocity
($\propto R^{-0.5}$) and gravitational force density is high.

We developed a simple model for the interaction of a periodic stellar wind
with an accreting environment. The model assumes spherical symmetry and
strong radiative losses from the wind and ambient shocks so that we can use a
thin shell approximation. The equations for mass and momentum conservation
for a shell of mass $M_{\rm s}$ and radius $R_{\rm s}$ are
\begin{equation} {d R_{\rm s} \over d t} = V_{\rm s} 
\label{ed1} 
\end{equation}
\begin{equation} {d M_{\rm s} \over d t} = 4 \pi {R_{\rm s}}^2 \rho_{\rm w}(R_{\rm s}) (V_{\rm w} -
V_{\rm s}) + 4 \pi {R_{\rm s}}^2 \rho_{\rm a}(R_{\rm s}) (V_{\rm s} + V_{\rm
		   a})
\label{ed2} 
\end{equation} 
\begin{equation} {d M_{\rm s} V_{\rm s} \over d t} = 4
\pi {R_{\rm s}}^2 \rho_{\rm w}(R_{\rm s}) V_{\rm w} (V_{\rm w} - V_{\rm s})
		   - 4 \pi {R_{\rm s}}^2 \rho_{\rm a}(R_{\rm s}) V_{\rm a}
		   (V_{\rm s} + V_{\rm a}) - {G M_* M_{\rm s} \over {R_{\rm
		   s}}^2}
\label{ed3} 
\end{equation} 
\cite{VoKw85}. Here $\rho_a$ and $V_a$ are the radially dependent accretion
flow density and velocity.  These equations can be rewritten in the form of a
simple set of coupled ODEs. We solved the ODEs using a 4th order Runge-Kutta
method with an adaptive step-size. For the variable wind we used a sinusoidal
variation with an amplitude of 100. $\dot M_{\rm w}$ is kept constant, but we
have also tried the case in which $\dot M_{\rm w}$ varies in the same way as
$V_{\rm w}$, which gives similar results.

The parameters for the wind and the environment are the same as in for the
model shown in Sect.~4.  Other details of the model can be found in Paper 2.
We assume that before the start of the integration, the bubble was set in
motion, perhaps by an energetic episodic outburst from the protostar. This
initial condition gives a total energy in the shell of $10^{40}$ ergs which
is more than five orders of magnitude less than what is released in a typical
FU Orionis outburst \cite{HartKen96}.

\begin{figure}
  \begin{center} \leavevmode
    \psfig{file=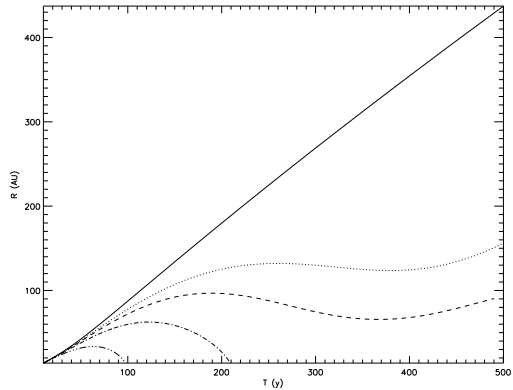,width=7cm,angle=90} \caption{Evolution
    of spherical wind blown bubbles in accretion environment.  Shown
    are the radius of 5 bubbles driven winds with different periods as
    a function of time.  Solid line corresponds to a period of $P =
    10^6$ years. Dotted line: $P = 500$ years. Dashed line: $P = 400$
    years.  Dashed-dot line: $P = 300$ years. Dashed-dot-dot line: $P
    = 200$ years.}\label{oscill}
  \end{center}
\end{figure}

The results for variable winds with periods of $\tau_{\rm w} = ~10^6$~, 500,
400, 300 and 200~years are shown in Fig.~\ref{oscill}.  For the constant wind
($\tau_{\rm w} = ~10^6$ ~years) the bubble expands monotonically although it
does experience changes in velocity due to accretion ram pressure and
gravitational forces.  When the wind is allowed to vary these forces produce
dramatic changes in the bubble's evolution.  For all four variation periods
we see that the expansion of bubble can be reversed ($V_{\rm s} < 0$) for
some time.  For longer periods the bubble gains enough momentum before the
wind enters a minimum to either continue a slow expansion ($\tau_{\rm w} =
500$ years) or maintain a constant average radius ($\tau_{\rm w} = 400$
years).  For shorter periods the bubble is ``crushed'' during the wind's
quiescent phase. Note that we ended the calculation if the shock radius
$R_{\rm s}$ became smaller than $10^{12}$~cm.

From these results we conclude that, in principle, time varying winds can
produce wind-blown bubbles whose size does not increase beyond some upper
limit.  If these bubbles produce jets through the hydrodynamic mechanisms
described above, the model age and collimation scales can be made consistent
with observations.  We imagine that during a periodic increase in mechanical
luminosity the YSO will begin driving a bubble which in turn produces
collimated jets.  As the wind speed varies the bubble either oscillates
around some average radius (producing variations in the jet beam) or it
collapses entirely. Jet production begins again when the momentum in the wind
has increased enough to produce another bubble.

\section{Connection to observations}
The results so far can be considered as physical experiments rather than
models for specific observed objects, since there are several important
effects which we have left out of the numerical simulations (gravity,
accretion flow, rotation). Clearly the next step should be to include these,
and also to add microphysics and diagnostics to make predictions of
observables, similarly to the work of
\citeauthor{FrMel94}~\shortcite{FrMel94} and
\citeauthor{Suttea97}~\shortcite{Suttea97}. This will allow us to check
whether the cool jets reproduce the observables in terms of temperature and
density and in how far the hot jet can be observed.

There are however already two observational results in support of the model.
Firstly the detection of non-thermal radio emission from the source W3(OH)
\cite{Reidea95}. These authors found strong synchrotron emission arising at
the center of a linear chain of maser sources. Recent high resolution images
show the synchrotron emission actually appears as a two sided jet emanating
from the geometric center of the two maser flows \cite{Wilnerea97}.  Such a
morphology is likely to occur from first-order Fermi acceleration at strong
shocks around the central object.

Secondly a geometry very reminiscent of the converging conical flows has been
observed in the proto-planetary nebulae He~3-1475 (Harrington, private
communication), an object which is producing collimated outflows with
velocities of several times 100~\kms\ \cite{Rieraea95}.

\section{Conclusions}
\begin{itemize}
\item The interaction of an outflow with a toroidal environment is an
efficient means to collimate unfocused or wide-angle winds.
\item The interaction gives rise to two jet phases, an initial `cool jet',
which form due to converging conical flows at the top of the bubble, and a
later `hot jet' which forms when cooling starts to be less efficient. The
cool jet is dense and is the one which should be most easily observed.
\item Both the variability and the longevity of the jets could be explained by
a model in which the outflow is time-dependent, producing a series of cool
jets.
\item The occurrence of non-thermal emission in some radio jets can be seen as
supporting our model, since its production requires strong shocks.
\end{itemize}


\vspace{0.2cm}

{\bf Acknowledgments:} The authors gratefully acknowledge Dong-Su Ryu
for making the TVD code available to us.  We also benefited from
discussions with Alberto Noriega-Crespo, Alex Raga, Mark Ried, Tom
Jones.  This work was supported by the Minnesota Supercomputer
Institute.

\end{document}